\documentclass[
twocolumn
 amsmath,amssymb,
 aps,
]{revtex4-2}

\usepackage{graphicx}
\usepackage{dcolumn}
\usepackage{bm}
\usepackage[colorlinks=true, allcolors=blue]{hyperref}

\usepackage{textcomp,graphicx,bm,amsmath,xcolor}
\usepackage{amssymb,color,caption,subcaption,setspace}

\usepackage{subcaption}

\begin{document}

\preprint{APS/123-QED}

\title{Metaharvesting: Emergent energy harvesting by piezoelectric metamaterials}

\author{Ibrahim Patrick\textsuperscript{1} }
\author{Sondipon Adhikari\textsuperscript{2} }
\author{Mahmoud I. Hussein\textsuperscript{3,4}}%
 \email{mih@colorado.edu}
\address{%
\textsuperscript{1}Bristol Composites Institute, Department of Aerospace Engineering, University of Bristol, Queen's Building, University Walk, Bristol, BS8 1TR, UK
\\
\textsuperscript{2}James Watt School of Engineering, The University of Glasgow, Glasgow, G12 8QQ, UK
\\
\textsuperscript{3}Ann and HJ Smead Department of Aerospace Engineering Sciences, University of Colorado Boulder, Boulder, Colorado 80303, USA
\\
\textsuperscript{4}Department of Physics, University of Colorado Boulder, Boulder, Colorado 80302, USA}%

\date{\today}

\begin{abstract} 
Vibration energy harvesting is a technology that enables electric power generation by augmenting vibrating materials or structures with piezoelectric elements. In a recent work, we quantified the intrinsic energy-harvesting availability of a piezoelectric phononic crystal (Piezo-PnC) by calculating its damping ratio across the Brillouin zone and subtracting off the damping ratio of the corresponding non-piezoelectric version of the phononic crystal. It was highlighted that the resulting quantity is indicative of the amount of useful energy available for harvesting and is independent of the finite structure size and boundary conditions and of any forcing conditions. Here we investigate the intrinsic energy harvesting availability of two other material systems chosen to be statically equivalent to a given Piezo-PnC: a piezoelectric locally resonant metamaterial (Piezo-LRM) and a piezoelectric inertially amplified metamaterial (Piezo-IAM). Upon comparing with the intrinsic energy harvesting availability of the Piezo-PnC, we observe an emergence of energy harvesting capacity, a phenomenon we refer to as metaharvesting. This is analogous to the concept of metadamping, except the quantity evaluated is associated with piezoelectric energy harvesting rather than raw dissipation. Our results show that the intrinsic energy harvesting availability is enhanced by local resonances, and enhanced further by inertial amplification. These findings open a pathway towards fundamental design of architectured piezoelectric materials with superior energy harvesting capacity.  
\end{abstract}

\maketitle


\section{Introduction}\label{section:introduction} 
Mechanical energy harvesting technologies aim to capture ambient energy, in various mechanical forms, and convert it into useful electrical energy~\cite{williams1996analysis, priya2009energy,elvin2013advances}. The advent of low-power energy-harvesting systems for driving devices that are usually battery operated provides new opportunities for the design of a multitude of energy-efficient products. Many engineering structures are susceptible to low-frequency structural vibrations that may be harnessed. A portion of the vibratory kinetic energy arising from ambient base-excited or forced mechanical vibrations may be captured rather than left to simply dissipate as loss. This process is facilitated by numerous approaches including the employment of piezoelectric materials~\cite{umeda1997energy}, shape memory alloys \cite{karaman2007energy}, ionic polymer metal composites \cite{aureli2009energy,cellini2014energy}, magnetostrictive materials \cite{wang2008vibration}, and structural nonlinearities \cite{mann2009energy,daqaq2010response,harne2013review,daqaq2014role}. Piezoelectric energy harvesting in particular has demonstrated promise for the practical generation of low-power electricity over broad frequency ranges~\cite{adhikari2009piezoelectric} and for a variety of applications including wireless sensors \cite{roundy2003energy}, micro-electrical devices \cite{beeby2006energy}, transducers \cite{priya2007advances}, micro-electromechanical-system (MEMS) portable devices \cite{cook2008powering}, structural health monitoring devices \cite{park2007structural,park2008energy,giurgiutiu2007structural}, biomedical implants \cite{amin2012powering}, among others. Multiple research reviews have presented in-depth overviews of advances in vibration energy harvesting technologies using piezoelectric elements added to engineering structures~\cite{sodano2004review,anton2007review,erturk2011piezoelectric,elvin2013advances,safaei2019review,sezer2021comprehensive}.


Another contemporary area of research that has also been rising rapidly over the past several decades is the study of elastic wave propagation in artificially structured  materials, widely referred to as \textit{phononic materials}~\cite{Mead1996,craster2012acoustic,deymier2013acoustic,hussein2014dynamics,ma2016acoustic,phani2017dynamics,laude2020phononic,jin2021physics}. In this domain, three key classes emerged: phononic crystals (PnC) \cite{nemat1972general,mead1975wave,sigalas1992elastic,Kushwaha1993,vasseur2001experimental,Phani2006,pennec2010two,willey2022coiled}, locally resonant acoustic/elastic metamaterials (LRM) \cite{liu2000locally,pennec2008low,wu2008evidence,Xiao2011formation,bastawrous2021theoretical}, and inertially amplified metamaterials (IAM)~\cite{yilmaz2007phononic,acar2013experimental,frandsen2016inertial,kulkarni2016longitudinal,banerjee2021inertial,jamil2022inerter}. In PnCs, the underlying dispersion curves are shaped by constructive and destructive interferences of waves that linearly scatter by periodic inclusions, interfaces, and/or boundaries within the medium~\cite{Kushwaha1993,laude2020phononic}.~In LRMs, which are often also periodic, the dispersion curves are shaped not only by wave interefences but also by couplings$-$or hybridizations$-$between substructure resonance modes and elastic wave modes in the hosting medium~\cite{liu2000locally,jin2021physics}. In contrast, in IAMs the unit cell is configured to allow inertial amplification to take place causing a magnification of the ``effective inertia" of a resonator.~This concept may be realized using a lever-arm effect that allows the inertia of a resonating mass to be magnified to a degree proportional to the arm length~\cite{yilmaz2007phononic,frandsen2016inertial} or using an~\textit{inerter}~\cite{smith2002synthesis} which is a component that creates inertial amplification by the rotation of an intrinsic flywheel~\cite{kulkarni2016longitudinal}. In either case, the inertial amplification component is repeated, usually periodically, to form an IAM with intrinsic wave propagation properties. In all these classes of materials, the concept of a band gap is a prominent feature since it represents a frequency range whereby wave propagation is spatially attenuated. Numerous research has explored approaches to optimize the unit-cell material and/or geometric properties to lower the frequencies of band gaps and maximize their widths~\cite{sigmund2003systematic,bilal2011ultrawide,badreddine2012enlargement,bilal2013trampoline}. In addition to band-gap spatial attenuation, temporal attenuation also takes place when damping is prescribed within the unit-cell constitution~\cite{hussein2009theory,hussein2010band}. For free wave motion, this is characterized by the wavenumber-dependent damping-ratio diagram, which is paired branch by branch with the frequency dispersion diagram. Both the frequency and damping ratio diagrams incorporate, in general, both real and imaginary wavenumbers~\cite{hussein2013microdynamics,Frazier_CRP_2016}. The level and frequency broadness of dissipative  temporal attenuation may be enhanced by the mechanism of \textit{metadamping}~\cite{hussein2013metadamping}. Metadamping emerges from the combination of local resonance~\cite{hussein2013metadamping,frazier2015viscous,bacquet2018dissipation,depauw2018metadamping,bacquet2018metadamping,aladwani2020mechanics} or inertial amplification~\cite{lin2021metadamping,hussein2022metadamping} with prescribed damping. With metadamping, the overall dissipation in relatively stiff unit cells may be designed to exceed the dissipation levels of the constituent components. Furthermore, this emergent property is realizable across relatively broad frequency ranges~\cite{bacquet2018dissipation}. The fundamental methodology for metadamping characterization is to compare the dissipation across the Brillouin zone for an elastic metamaterial (LRM or IAM) and a statically equivalent phononic crystal, or conventional material~\cite{bacquet2018dissipation}, with the same level of overall prescribed damping. The difference in the damping ratio between the two systems represents the quantity of metadamping, which may be positive as demonstrated by Hussein and Frazier~\cite{hussein2013metadamping} and in subsequent studies or negative to indicate inhibited loss~\cite{frazier2015viscous,bacquet2018dissipation, depauw2018metadamping,bacquet2018metadamping,aladwani2020mechanics,lin2021metadamping,hussein2022metadamping}. A key advantage of metadamping is that it allows for~\textit{dissipation engineering} in a manner that is effectively independent of the material's overall stiffness~\cite{hussein2013metadamping}. 
\begin{figure}[b]
\centering
\includegraphics{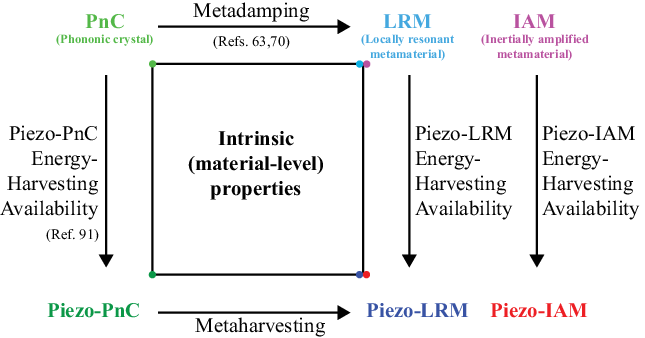}
\caption{Schematic of an "analysis square" highlighting key intrinsic property connections between statically equivalent PnC, LRM/IAM, Piezo-PnC, and Piezo-LRM/Piezo-IAM, each located at a corner. All models have the same type and level of prescribed damping. Horizontal connections indicate metadamping (raw dissipation characterizations, top) or metaharvesting (piezoelectric characterizations, bottom), and vertical connections indicate EHA characterization for the three phononic materials considered.}
\label{fig01}
\end{figure}
\begin{figure}[t]
\centering
\includegraphics{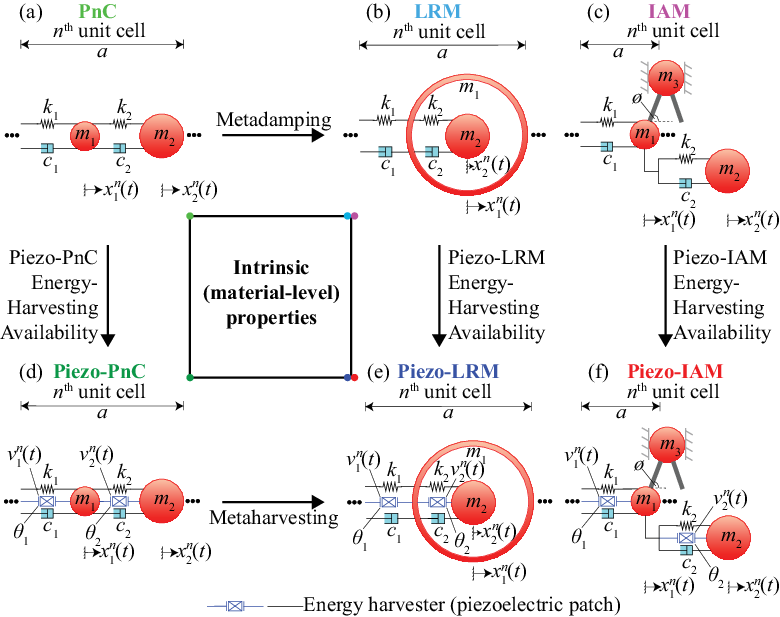}
\caption{Unit-cell schematics of statically equivalent (a) PnC, (b) LRM, (c) IAM, (d) Piezo-PnC, (e) Piezo-LRM, and (f) Piezo-IMA. All models have the same number and type of damping dashpots with the same values of prescribed damping. Arrows indicating metadamping, EHA, and metaharvesting characterization are included in the appropriate locations.}\label{fig:unit_cell_schematics}
\end{figure}
The merging of the two aforementioned fields$-$piezoelectric energy harvesting and phononic materials$-$has subsequently evolved as a natural development. Incorporation of piezoelectric components into PnC architecture for active elastodynamic tuning or vibration energy harvesting has been the focus of numerous studies~\cite{thorp2001attenuation,hou2004phononic,laude2005full,gonella2009interplay,rupp2010switchable,chen2013broadband,park2019two,cao2019vibration,shin2020phononic}. Similarly, piezoelectric tuning was applied to LRMs \cite{Airoldi_NJP_2011,casadei2012piezoelectric,hu2017metastructure,li2017design,hu2018internally} and piezoelectric energy harvesting was realized using LRMs~\cite{mikoshiba2013energy,shen2015low,hu2017metastructure,li2017design,hu2018internally}. The reader is referred to Lee \textit{et al.} \cite{lee2022piezoelectric}, Hu \textit{et al.} \cite{hu2021acoustic}, and Chen \textit{et al.} \cite{chen2014metamaterials} for review articles on piezoelectric energy harvesting using PnCs and/or LRMs. In the same spirit, IAMs have been linked with energy harvesting where a configuration comprising a finite cantilever beam with bimorph piezoelectric layers and an inertially amplified tip mass was investigated~\cite{adhikari2021enhanced}.


The common approach for characterizing the energy harvesting capacity in both conventional and phononic-material-based systems is to do so at the extrinsic \textit{structural} level, i.e., with considerations pertaining to overall size, boundary conditions, and forcing information. In a previous work~\cite{patrick2021brillouin}, we presented an alternative approach whereby the characterization is done at the intrinsic \textit{material} level.\footnote{The reader is referred to Refs.~\cite{Al-Babaa_2017_Poles,Bastawrous_2022,al2023theory,rosa2023material} for a selection of recent studies on the effects of truncating a phononic material to form a finite phononic structure.} Specifically, a PnC with integrated piezoelectric elements was considered and analysed by Bloch's theorem as a damped medium~\cite{hussein2009theory,hussein2010band}. In addition to the frequency dispersion curves, the wavenumber-dependent damping ratio curves (which provide a measure of dissipation capacity for free wave motion) were obtained and compared directly with the corresponding set for the same PnC but without the piezoelectric elements. The dissipation curves for the latter, the non-piezoelectric PnC, was shown to indicate the ‘raw’ dissipation representing unutilized/lost energy. On the other hand, the difference
in the damping ratio curves between the two models$-$the PnC with piezoelectric elements (i.e., Piezo-PnC) versus the PnC without piezoelectric elements$-$was shown to offer a formal wavenumber-dependent representation of the amount of \textit{energy harvesting availability} (EHA). This Brillouin zone-
based approach for energy-harvesting characterization follows the characterization framework for metadamping~\cite{hussein2013metadamping}, except the quantity of interest is the amount of "dissipation" available for harvesting as opposed to raw dissipation.

In this paper we characterize the EHA of a piezoelectric LRM (Piezo-LRM) and a piezoelectric IAM (Piezo-IAM), both selected to exhibit identical static (i.e., long-wave) and prescribed damping properties to a demonstrative model of a diatomic Piezo-PnC. For all piezoelectric models, only shunted circuits with an inductor are considered for the piezoelectric elements. We then compare the EHA of the Piezo-LRM and Piezo-IAM with that of the Piezo-PnC. By substracting the EHA of the Piezo-PnC from that of the Piezo-LRM or Piezo-IAM, we generate a formal quantity that describes the emergent enhancement of energy harvesting capacity, which we denote \textit{metaharvesting}. For completion, we also characterize the metadamping properties of the non-piezoelectric LRM and IAM compared to the non-piezoelectric PnC. In doing so, we complete an "analysis square" that has PnC and LRM/IAM on the top left and right corners respectively, and Piezo-PnC and Piezo-LRM/Piezo-IAM on the bottom left and right corners respectively. Thus the top side represents the property of metadamping, with an arrow pointing from left to right, and the bottom side represent the property of metaharvesting with an arrow also pointing from left to right. In analogy, the left side of the square represents Piezo-PnC EHA with an arrow going from top to bottom, and, similarly, the right side of the square represents Piezo-LRM or Piezo-IAM with arrows going from top to bottom. This four-way analysis framework is laid out in Fig. ~\ref{fig01}. A key advantage of these characterizations is that they are all intrinsic in nature. Another inherent advantage is that by necessity of the characterization protocol all six models considered are statically equivalent with the same level of prescribed damping, and are therefore directly comparable with each other. This unified approach enables a single damping ratio-versus-frequency characterization diagram for all models at once as illustrated later in the paper.  


The layout of the paper is as follows. Section~\ref{section:Bloch_analysis} details the governing equations and the application of Bloch's theorem to the unit cells of both the non-piezoelectric and piezoelectric periodic materials considered. In Section~\ref{section:wave_propg_results}, a comparative analysis is provided, aided with a systematic set of characterization plots of the dispersion and dissipation characteristics of the non-piezoelectric and piezoelectric phononic materials, addressing each of the arrows in Fig.~\ref{fig01}. In Section~\ref{section:summary_conclusions}, we present the unified characterization diagram mentioned above, along with concluding remarks. 

\section{Bloch analysis of non-piezoelectric and piezoelectric phononic materials}\label{section:Bloch_analysis}

In this section, we introduce all six of the phononic materials we consider in this investigation, first without and then with the piezoelectric elements incorporated. Each phononic material is represented by its unit-cell schematic and examined using a lumped-parameter model as shown in Fig.~\ref{fig:unit_cell_schematics}. In the figure, $a$ is the length of the lattice unit cell, $\phi$ is the angle between the central axis of the IAM and the rigid links of the inertial-amplifier attachment; $m_1$ and $m_2$ denote the values of the masses, $k_1$ and $k_2$ are the stiffness coefficients of the springs, and $c_1$ and $c_2$ are the damping coefficients of the viscous damping dashpots; $\theta_1$ and $\theta_2$ denote the electromechanical couplings of the energy harvesters; $n$ is used to identify the central unit cell under consideration; $x_1^{n}(t)$ and $x_2^{n}(t)$ are the displacements of the masses, as a function of time $t$, in the $n^{\textrm{th}}$ unit cell; and $v_1^{n}(t)$ and $v_2^{n}(t)$ are the voltages for the corresponding masses. Note that the IAM unit cell is similar to the LRM unit cell except that in addition to each baseline mass connected to a local resonator, each baseline mass is connected to the neighbouring baseline masses by inertial-amplifier attachments, each of which is comprised of an auxiliary mass $m_3$ and two rigid links, one for each of the neighbouring baseline masses.

\subsection{Phononic crystal, locally resonant metamaterial, and inertially amplified metamaterial without piezoelectric elements}\label{subsection:non_piezo_periodic_media}

\subsubsection{Phononic crystal}\label{subsubsection:PnC}

By applying a force balance on the masses, the governing equations pertaining to the $n^{\textrm{th}}$ unit cell of the diatomic PnC shown in Fig.~\ref{fig:unit_cell_schematics}(a) are written as
\begin{align}
m_1\ddot{x}_1^n+(c_1+c_2)\dot{x}_1^n-c_2\dot{x}_2^n-c_1\dot{x}_2^{n-1}+(k_1+k_2)x_1^n-k_2x_2^n-k_1x_2^{n-1}&=0,\label{eq:PnC_gov_eq_1}\\
m_2\ddot{x}_2^n+(c_1+c_2)\dot{x}_2^n-c_2\dot{x}_1^n-c_1\dot{x}_1^{n+1}+(k_1+k_2)x_2^n-k_2x_1^n-k_1x_1^{n+1}&=0,\label{eq:PnC_gov_eq_2}
\end{align}
where the number of overhead dots indicates the order of the derivative with respect to time, and $(n+1)$ and $(n-1)$  refer to the unit cells to the right and left of the central ($n^{\textrm{th}}$) unit cell, respectively. 

Applying Bloch's theorem~\cite{hussein2009theory,hussein2010band,hussein2014dynamics}, the displacements $x_l$ of the masses are given by
\begin{equation}\label{eq:Bloch_solution_displacement}
x_l^{n+g}(r,\kappa;t)=\tilde{x}_{l}(t)\textrm{e}^{\textrm{i}\kappa(n+g)a},
\end{equation}
where $l=1,2$ is an index for each mass in a unit cell, $g\in[-\infty,\infty]$ is an integer used to locate and refer to any unit cell relative to the central unit cell under consideration, i.e., the $n^{\textrm{th}}$ unit cell shown in Fig.~\ref{fig:unit_cell_schematics}, $r$ is the one-dimensional position vector of the $n^{\textrm{th}}$ unit cell given by $r=na$, $\kappa$ is the wavenumber, $\tilde{x}_{l}$ is the displacement wave amplitude, and $\textrm{i}=\sqrt{-1}$ is the imaginary unit. For the central unit cell under consideration, $g=0$ ($n^{\textrm{th}}$ unit cell), and for the unit cells to the left and right of the central unit cell, $g=-1$, and $g=+1$, respectively; i.e., the $(n-1)^{\textrm{th}}$ and $(n+1)^{\textrm{th}}$ unit cells, respectively. Substituting Eq. (\ref{eq:Bloch_solution_displacement}) into Eqs. (\ref{eq:PnC_gov_eq_1}) and (\ref{eq:PnC_gov_eq_2}) yields two Bloch transformed equations for the displacement amplitudes, $\tilde{x}_1$ and $\tilde{x}_2$, which can be written in a matrix form as
\begin{equation}\label{eq:PnC_bloch_matrix_eq}
\mathbf{M}\mathbf{\ddot{\tilde{X}}}+\mathbf{C}(\kappa)\mathbf{\dot{\tilde{X}}}+\mathbf{K}(\kappa)\mathbf{\tilde{X}}=\mathbf{0},
\end{equation}
where
\begin{align}\label{eq:PnC_bloch_matrices}
\mathbf{\tilde{X}}=\begin{pmatrix}\tilde{x}_{1}\\\tilde{x}_{2}\end{pmatrix},\ \mathbf{M}=\begin{pmatrix}m_1&0\\0&m_2\end{pmatrix},\ \mathbf{C}(\kappa)=\begin{pmatrix}c_1+c_2&-c_2-c_1\textrm{e}^{-\textrm{i}\kappa a}\\-c_2-c_1\textrm{e}^{\textrm{i}\kappa a}&c_1+c_2\end{pmatrix},\cr 
\textrm{and }\mathbf{K}(\kappa)=\begin{pmatrix}k_1+k_2&-k_2-k_1\textrm{e}^{-\textrm{i}\kappa a}\\-k_2-k_1\textrm{e}^{\textrm{i}\kappa a}&k_1+k_2\end{pmatrix}.
\end{align}
By means of state-space transformation, Eq. (\ref{eq:PnC_bloch_matrix_eq}) is converted into a first-order state-space equation of the form
\begin{equation}\label{eq:state_space_eq}
\mathbf{A}\mathbf{\dot{Y}}+\mathbf{B}\mathbf{Y}=\mathbf{0},
\end{equation}
where
\begin{equation}\label{eq:PnC_state_space_matrices}
\mathbf{A}=\begin{pmatrix}\mathbf{0}&\mathbf{I}\\\mathbf{M}&\mathbf{C}(\kappa)\end{pmatrix},\ \mathbf{B}=\begin{pmatrix}\mathbf{-I}&\mathbf{0}\\\mathbf{0}&\mathbf{K}(\kappa)\end{pmatrix},\textrm{ and } \mathbf{Y}=\begin{pmatrix}\mathbf{\dot{\tilde{X}}}\\\mathbf{\tilde{X}}\end{pmatrix}.
\end{equation}
For Eq. (\ref{eq:state_space_eq}), a solution of the form $\mathbf{Y}=\mathbf{\tilde{Y}}_\lambda\textrm{e}^{\lambda t}$ is assumed, where $\mathbf{\tilde{Y}}_\lambda$ is a complex-amplitude state-space vector corresponding to eigenvalue $\lambda$. The dispersion relation is obtained by solving Eq. (\ref{eq:state_space_eq}) as an eigenvalue problem given by 
\begin{equation}\label{eq:eigenvalue_problem}
\left\vert\mathbf{A}^{-1}\mathbf{B}+\lambda\mathbf{I}\right\vert=0.
\end{equation}
Expanding Eq. (\ref{eq:eigenvalue_problem}) yields a fourth-order equation in terms of $\lambda$, which, upon solving, gives four complex roots appearing as two complex-conjugate pairs. The complex solution for the eigenvalue problem at a given value of $\kappa$ is expressed as 
\begin{equation}\label{eq:eigenvalue_solution}
\lambda_l(\kappa)=-\zeta_l(\kappa)\omega_{\textrm{r}_l}(\kappa)\pm \textrm{i}\omega_{\textrm{d}_l}(\kappa)=-\zeta_l(\kappa)\omega_{\textrm{r}_l}(\kappa)\pm \textrm{i}\omega_{\textrm{r}_l}(\kappa)\sqrt{1-\zeta_l(\kappa)^2},
\end{equation}
where the subscript $l$ refers to each complex-conjugate pair and, consequently, the mode or branch number. Considering that all the unit cells shown in Fig. \ref{fig:unit_cell_schematics} have two degrees of freedom and their complex eigen solution comprises two complex-conjugate pairs, they exhibit two modes or branches: the acoustic (lower/first) branch given by $l=1$ and the optical (higher/second) branch given by $l=2$. The definitions for the wavenumber-dependent resonant frequency $\omega_{\textrm{r}_l}$, damped frequency $\omega_{\textrm{d}_l}$, and damping ratio $\zeta_l$ can be extracted from Eq. (\ref{eq:eigenvalue_solution}) in the following manner:   
\begin{align}
\omega_{\textrm{r}_l}(\kappa)&=\textrm{Abs}[\lambda_l(\kappa)],\label{eq:resonant_freq_solution}\\
\omega_{\textrm{d}_l}(\kappa)&=\textrm{Im}[\lambda_l(\kappa)],\label{eq:damped_freq_solution}\\
\zeta_l(\kappa)&=-\frac{\textrm{Re}[\lambda_l(\kappa)]}{\textrm{Abs}[\lambda_l(\kappa)]}.\label{eq:damping_ratio_solution}
\end{align}

\subsubsection{Locally resonant metamaterial}\label{subsubsection:LRM}

A force balance on the $n^{\textrm{th}}$ unit cell of the LRM model shown in Fig.~\ref{fig:unit_cell_schematics}(b) yields the following governing equations 
\begin{align}
m_1\ddot{x}_1^n+c_1(2\dot{x}_1^n-\dot{x}_1^{n-1}-\dot{x}_1^{n+1})+c_2(\dot{x}_1^n-\dot{x}_2^n)+k_1(2x_1^n-x_1^{n-1}-x_1^{n+1})\cr
+k_2(x_1^n-x_2^n)=0,\label{eq:LRM_gov_eq_1}\\
m_2\ddot{x}_2^n+c_2(\dot{x}_2^n-\dot{x}_1^n)+k_2(x_2^n-x_1^n)=0.\label{eq:LRM_gov_eq_2}
\end{align}

Following Bloch transformation of Eqs. (\ref{eq:LRM_gov_eq_1}) and (\ref{eq:LRM_gov_eq_2}), the dispersion relation of the LRM is obtained in a similar manner as for the PnC in Section 2(a)(\ref{subsubsection:PnC}) leading to Eq. (\ref{eq:PnC_bloch_matrix_eq}) where
the $\mathbf{C(\kappa)}$ and $\mathbf{K(\kappa)}$ matrices are 
\begin{align}\label{eq:LRM_bloch_matrices}
\mathbf{C}(\kappa)=\begin{pmatrix}c_1(2-\textrm{e}^{-\textrm{i}\kappa a}-\textrm{e}^{\textrm{i}\kappa a})+c_2 & -c_2 \\ -c_2 & c_2\end{pmatrix},\cr
\textrm{and }\mathbf{K}(\kappa)=\begin{pmatrix}k_1(2-\textrm{e}^{-\textrm{i}\kappa a}-\textrm{e}^{\textrm{i}\kappa a})+k_2 & -k_2 \\ -k_2 & k_2\end{pmatrix},
\end{align} 
and where $\mathbf{\tilde{X}}$ and $\mathbf{M}$ are the same as in Eq. (\ref{eq:PnC_bloch_matrices}).

\subsubsection{Inertially amplified metamaterial}\label{subsubsection:IALRM}

In the IAM unit-cell model shown in Fig.~\ref{fig:unit_cell_schematics}(c), the inertial-amplifier (auxiliary) mass $m_3$ is connected to the baseline mass $m_1$ by a rigid link. Thus the inertial rigid coupling does not alter the total degrees of freedom of a unit cell and the IAM unit cell will exhibit only two degrees of freedom similar to the PnC and LRM. The governing equations pertaining to the $n^{\textrm{th}}$ unit cell of the IAM are
\begin{align}
m_1\ddot{x}_1^n+c_1(2\dot{x}_1^n-\dot{x}_1^{n-1}-\dot{x}_1^{n+1})+c_2(\dot{x}_1^n-\dot{x}_2^n)+k_1(2x_1^n-x_1^{n-1}-x_1^{n+1})\cr
+k_2(x_1^n-x_2^n)+\chi(m_3(\ddot{x}_1^n-\ddot{x}_1^{n-1}))-\chi(m_3(\ddot{x}_1^{n+1}-\ddot{x}_1^n))&=0,\label{eq:IALRM_gov_eq_1}\\
m_2\ddot{x}_2^n+c_2(\dot{x}_2^n-\dot{x}_1^n)+k_2(x_2^n-x_1^n)&=0,\label{eq:IALRM_gov_eq_2}
\end{align}
where $\chi=1/(4\tan^2\phi)$ as dictated from the kinematics of the model.

Following a Bloch transformation of equations (\ref{eq:IALRM_gov_eq_1}) and (\ref{eq:IALRM_gov_eq_2}), the dispersion relation of the IALRM is obtained in a manner similar as for the PnC in section \ref{subsubsection:PnC}. In equation (\ref{eq:PnC_bloch_matrix_eq}) for the IALRM, $\mathbf{\tilde{X}}$ is the same as in equation (\ref{eq:PnC_bloch_matrices}); $\mathbf{C}(\kappa)$ and $\mathbf{K}(\kappa)$ are the same as in equation (\ref{eq:LRM_bloch_matrices}); and
\begin{equation}\label{eq:IALRM_block_matrix}
\mathbf{M}=\begin{pmatrix}m_1+\displaystyle \frac{m_3}{4}(2-\textrm{e}^{-\textrm{i}\kappa a}-\textrm{e}^{\textrm{i}\kappa a})\cot^2\phi & 0 \\ 0 & m_2\end{pmatrix}.
\end{equation}

\subsection{Phononic crystal, locally resonant metamaterial, and inertially amplified metamaterial with shunted piezoelectric elements}\label{subsection:piezo_periodic_media}

\subsubsection{Piezoelectric phononic crystal}\label{subsubsection:PPnC}

The governing electromechanical equations for the $n^{\textrm{th}}$ unit cell of the Piezo-PnC shown in Fig.~\ref{fig:unit_cell_schematics}(c) are written as
\begin{align}
m_1\ddot{x}_1^n+(c_1+c_2)\dot{x}_1^n-c_2\dot{x}_2^n-c_1\dot{x}_2^{n-1}+(k_1+k_2)x_1^n-k_2x_2^n-k_1x_2^{n-1}\cr
+\theta_1v_1^n-\theta_2v_2^n&=0,\label{eq:PPnC_gov_eq_1}\\
m_2\ddot{x}_2^n+(c_1+c_2)\dot{x}_2^n-c_2\dot{x}_1^n-c_1\dot{x}_1^{n+1}+(k_1+k_2)x_2^n-k_2x_1^n-k_1x_1^{n+1}\cr
+\theta_2v_2^n-\theta_1v_1^{n+1}&=0,\label{eq:PPnC_gov_eq_2}\\
-\theta_1(\ddot{x}_1^n-\ddot{x}_2^{n-1})+C_{p_{1}}\ddot{v}_1^n+\frac{1}{R_1}\dot{v}_1^n+\frac{1}{L_1}v_1^n&=0,\label{eq:PPnC_gov_eq_3}\\
-\theta_2(\ddot{x}_2^n-\ddot{x}_1^n)+C_{p_{2}}\ddot{v}_2^n+\frac{1}{R_2}\dot{v}_2^n+\frac{1}{L_2}v_2^n&=0.\label{eq:PPnC_gov_eq_4}
\end{align}
In Eqs. (\ref{eq:PPnC_gov_eq_3}) and (\ref{eq:PPnC_gov_eq_4}), $C_{p_l}$, $R_l$, and $L_l$, where $l=1,2$, represent the capacitance, resistance, and inductance associated with the shunt circuits of the piezoelectric element for each degree of freedom, respectively. 

Similar to the non-piezoelectric models,  Bloch's theorem is applied to the corresponding piezoelectric models~\cite{patrick2021brillouin}. It is applied to the motion variable as in Eq.~(\ref{eq:Bloch_solution_displacement}), and it is similarly applied to the piezoelectric element voltage variable in the unit cell,
\begin{equation}\label{eq:Bloch_solution_voltage}
v_l^{n+g}(r,\kappa;t)=\tilde{v}_l(t)\textrm{e}^{\textrm{i}\kappa(n+g)a},
\end{equation}
where $\tilde{v}$ is the voltage amplitude. Substituting Eqs. (\ref{eq:Bloch_solution_displacement}) and (\ref{eq:Bloch_solution_voltage}) into Eqs. (\ref{eq:PPnC_gov_eq_1})--(\ref{eq:PPnC_gov_eq_4}) yields four homogeneous equations, two for the displacement amplitudes $\tilde{x}_{1}$ and $\tilde{x}_{2}$ and two for the voltage amplitudes $\tilde{v}_{1}$ and $\tilde{v}_{2}$, which are written in matrix form as
\begin{align}
\mathbf{M}\mathbf{\ddot{\tilde{X}}}+\mathbf{C}(\kappa)\mathbf{\dot{\tilde{X}}}+\mathbf{K}(\kappa)\mathbf{\tilde{X}}+\mathbf{T_1}(\kappa)\mathbf{\tilde{V}}&=\mathbf{0},\label{eq:PPnC_bloch_matrix_eq_1}\\
\mathbf{T_2}(\kappa)\mathbf{\ddot{\tilde{X}}}+\mathbf{C_p}\mathbf{\ddot{\tilde{V}}}+\mathbf{R}\mathbf{\dot{\tilde{V}}}+\mathbf{L}\mathbf{\tilde{V}}&=\mathbf{0}.\label{eq:PPnC_bloch_matrix_eq_2}
\end{align}
In Eqs.~(\ref{eq:PPnC_bloch_matrix_eq_1}) and (\ref{eq:PPnC_bloch_matrix_eq_2}), $\mathbf{\tilde{X}}$, $\mathbf{M}$, $\mathbf{C}(\kappa)$, and $\mathbf{K}(\kappa)$ are the same as in equation (\ref{eq:PnC_bloch_matrices}) and the remaining vectors and matrices are defined as
\begin{align}\label{eq:PPnC_bloch_matrices}
\mathbf{\tilde{V}}=\begin{pmatrix}\tilde{v}_{1}\\\tilde{v}_{2}\end{pmatrix},\ \ \mathbf{T_1}(\kappa)=\begin{pmatrix}\theta_1&-\theta_2\\-\theta_1\textrm{e}^{\textrm{i}\kappa a}&\theta_2\end{pmatrix},\ \ \mathbf{T_2}(\kappa)=\begin{pmatrix}-\theta_1&\theta_1\textrm{e}^{-\textrm{i}\kappa a}\\\theta_2&-\theta_2\end{pmatrix},\cr
\mathbf{C_p}=\begin{pmatrix}C_{p_1}&0\\0&C_{p_2}\end{pmatrix},\ \  \mathbf{R}=\begin{pmatrix}\displaystyle\frac{1}{R_1}&0\\0&\displaystyle\frac{1}{R_2}\end{pmatrix},\textrm{ and }\mathbf{L}=\begin{pmatrix}\displaystyle\frac{1}{L_1}&0\\0&\displaystyle\frac{1}{L_2}\end{pmatrix}.
\end{align}
Equations (\ref{eq:PPnC_bloch_matrix_eq_1}) and (\ref{eq:PPnC_bloch_matrix_eq_2}) can be compacted into a single matrix equation as
\begin{equation}\label{eq:PPnC_compact_bloch_matrix_eq}
\mathbf{Z_1}\mathbf{\ddot{\tilde{E}}}+\mathbf{Z_2}\mathbf{\dot{\tilde{E}}}+\mathbf{Z_3}\mathbf{\tilde{E}}=\mathbf{0},
\end{equation}
where
\begin{align}\label{eq:PPnC_compact_bloch_matrices}
\mathbf{Z_1}=\begin{pmatrix}\mathbf{M}&\mathbf{0}\\\mathbf{T_2}(\kappa)&\mathbf{C_p}\end{pmatrix},\ \mathbf{Z_2}=\begin{pmatrix}\mathbf{C}(\kappa)&\mathbf{0}\\\mathbf{0}&\mathbf{R}\end{pmatrix},\ \mathbf{Z_3}=\begin{pmatrix}\mathbf{K}(\kappa)&\mathbf{T_1}(\kappa)\\\mathbf{0}&\mathbf{L}\end{pmatrix},\cr
\textrm{and } \mathbf{\tilde{E}}=\begin{pmatrix}\mathbf{\tilde{X}}\\\mathbf{\tilde{V}}\end{pmatrix}.
\end{align}
The dispersion relation can now be formulated by subjecting Eq.~(\ref{eq:PPnC_compact_bloch_matrix_eq}) to a state-space transformation of the form given in equation (\ref{eq:state_space_eq}), where
\begin{equation}\label{eq:PPnC_state_space_matrices}
\mathbf{A}=\begin{pmatrix}\mathbf{0}&\mathbf{I}\\\mathbf{Z_1}&\mathbf{Z_2}\end{pmatrix},\ \mathbf{B}=\begin{pmatrix}\mathbf{-I}&\mathbf{0}\\\mathbf{0}&\mathbf{Z_3}\end{pmatrix},\textrm{ and }\mathbf{Y}=\begin{pmatrix}\mathbf{\dot{\tilde{E}}}\\\mathbf{\tilde{E}}\end{pmatrix},
\end{equation}
for which a solution of the form $\mathbf{Y}=\mathbf{\tilde{Y}}_\lambda\textrm{e}^{\lambda t}$ is assumed and a corresponding  eigenvalue problem of the form given in Eq.~(\ref{eq:eigenvalue_problem}) is obtained. Expanding Eq.~(\ref{eq:eigenvalue_problem}) for the Piezo-PnC model yields an eight-order equation in terms of $\lambda$, which, upon solving, gives four complex roots appearing as two complex-conjugate pairs and four real roots.

\subsubsection{Piezoelectric locally resonant metamaterial}\label{subsubsection:LRPM}

The governing electromechanical equations pertaining to the $n^{\textrm{th}}$ unit cell of the Piezo-LRM depicted in Fig.~\ref{fig:unit_cell_schematics}(e) is written as
\begin{align}
m_1\ddot{x}_1^n+c_1(2\dot{x}_1^n-\dot{x}_1^{n-1}-\dot{x}_1^{n+1})+c_2(\dot{x}_1^n-\dot{x}_2^n)+k_1(2x_1^n-x_1^{n-1}-x_1^{n+1})\cr
+k_2(x_1^n-x_2^n)+\theta_1v_1^n-\theta_1v_1^{n+1}-\theta_2v_2^n&=0,\label{eq:LRPM_gov_eq_1}\\
m_2\ddot{x}_2^n+c_2(\dot{x}_2^n-\dot{x}_1^n)+k_2(x_2^n-x_1^n)+\theta_2v_2^n&=0,\label{eq:LRPM_gov_eq_2}\\
-\theta_1(\ddot{x}_1^n-\ddot{x}_1^{n-1})+C_{p_{1}}\ddot{v}_1^n+\frac{1}{R_1}\dot{v}_1^n+\frac{1}{L_1}v_1^n&=0,\label{eq:LRPM_gov_eq_3}\\
-\theta_2(\ddot{x}_2^n-\ddot{x}_1^n)+C_{p_{2}}\ddot{v}_2^n+\frac{1}{R_2}\dot{v}_2^n+\frac{1}{L_2}v_2^n&=0.\label{eq:LRPM_gov_eq_4}
\end{align}

Following a Bloch transformation of Eqs. (\ref{eq:LRPM_gov_eq_1})--(\ref{eq:LRPM_gov_eq_4}), the dispersion relation for the Piezo-LRM is obtained in a similar manner as for the Piezo-PnC in Section 2(b)(\ref{subsubsection:PPnC}) yielding Eqs. (\ref{eq:PPnC_bloch_matrix_eq_1}) and (\ref{eq:PPnC_bloch_matrix_eq_2}). For  the Piezo-LRM, $\mathbf{\tilde{X}}$ and $\mathbf{M}$ are the same as in Eq.~(\ref{eq:PnC_bloch_matrices}), $\mathbf{C}(\kappa)$ and $\mathbf{K}(\kappa)$ are the same as in Eq.~(\ref{eq:LRM_bloch_matrices}),  $\mathbf{\tilde{V}}$, $\mathbf{C_p}$, $\mathbf{R}$, and $\mathbf{L}$ are the same as in Eq.~(\ref{eq:PPnC_bloch_matrices}), and $\mathbf{T_1}$ and $\mathbf{T_2}$ are defined as
\begin{equation}\label{eq:LRPM_bloch_matrices}
\mathbf{T_1}(\kappa)=\begin{pmatrix}\theta_1-\theta_1\textrm{e}^{\textrm{i}\kappa a}&-\theta_2\\0&\theta_2\end{pmatrix}\textrm{ and }\mathbf{T_2}(\kappa)=\begin{pmatrix}-\theta_1+\theta_1\textrm{e}^{-\textrm{i}\kappa a}&0\\\theta_2&-\theta_2\end{pmatrix}.
\end{equation}

\subsubsection{Piezoelectric inertially amplified metamaterial}\label{subsubsection:IAPM}

The governing electromechanical equations for the $n^{\textrm{th}}$ unit cell of the Piezo-IAM presented in Fig.~\ref{fig:unit_cell_schematics}(f) are
\begin{align}
m_1\ddot{x}_1^n+\chi(m_3(\ddot{x}_1^n-\ddot{x}_1^{n-1}))-\chi(m_3(\ddot{x}_1^{n+1}-\ddot{x}_1^n))+c_1(2\dot{x}_1^n-\dot{x}_1^{n-1}-\dot{x}_1^{n+1})\cr
+c_2(\dot{x}_1^n-\dot{x}_2^n)+k_1(2x_1^n-x_1^{n-1}-x_1^{n+1})+k_2(x_1^n-x_2^n)+\theta_1v_1^n-\theta_1v_1^{n+1}-\theta_2v_2^n&=0,\label{eq:IALRPM_gov_eq_1}\\
m_2\ddot{x}_2^n+c_2(\dot{x}_2^n-\dot{x}_1^n)+k_2(x_2^n-x_1^n)+\theta_2v_2^n&=0,\label{eq:IALRPM_gov_eq_2}\\
-\theta_1(\ddot{x}_1^n-\ddot{x}_1^{n-1})+C_{p_{1}}\ddot{v}_1^n+\frac{1}{R_1}\dot{v}_1^n+\frac{1}{L_1}v_1^n&=0,\label{eq:IALRPM_gov_eq_3}\\
-\theta_2(\ddot{x}_2^n-\ddot{x}_1^n)+C_{p_{2}}\ddot{v}_2^n+\frac{1}{R_2}\dot{v}_2^n+\frac{1}{L_2}v_2^n&=0.\label{eq:IALRPM_gov_eq_4}
\end{align}

Applying Bloch's theorem to Eqs.~(\ref{eq:IALRPM_gov_eq_1})--(\ref{eq:IALRPM_gov_eq_4}), the dispersion relation for the Piezo-IAM is obtained in a manner similar as for the Piezo-PnC in Section 2(b)(\ref{subsubsection:PPnC}). In Eqs.~(\ref{eq:PPnC_bloch_matrix_eq_1}) and (\ref{eq:PPnC_bloch_matrix_eq_2}) for the Piezo-IAM, $\mathbf{\tilde{X}}$ is the same as in (\ref{eq:PnC_bloch_matrices}), $\mathbf{M}$ is the same as in Eq.(\ref{eq:IALRM_block_matrix}), $\mathbf{C}(\kappa)$ and $\mathbf{K}(\kappa)$ are the same as in Eq.~(\ref{eq:LRM_bloch_matrices}), $\mathbf{T_1}(\kappa)$ and $\mathbf{T_2}(\kappa)$ are the same as in Eq.~(\ref{eq:LRPM_bloch_matrices}), and $\mathbf{\tilde{V}}$, $\mathbf{C_p}$, $\mathbf{R}$, and $\mathbf{L}$ are the same as in Eq.~ (\ref{eq:PPnC_bloch_matrices}).

\section{Wave-propagation and dissipation characteristics: Metadamping, energy-harvesting availability, and metaharvesting}\label{section:wave_propg_results}

In this section, the mathematical realizations of metadamping~\cite{hussein2013metadamping} and EHA~\cite{patrick2021brillouin} are first reviewed and the mathematical definition of metaharvesting is then presented. In this manner, each of the arrows shown in Fig.~\ref{fig:unit_cell_schematics} is addressed by comparing the dissipation characteristics of the various periodic media, with the  parameters for all models listed in Table~\ref{table:parameters}. 
\begin{table}[!h]
\caption{Mechanical and non-dimensional electrical parameters considered in all unit cells investigated, namely, PnC, Piezo PnC, LRM, Piezo LRM, IAM and Piezo IAM; the mechanical parameters are the same for the non-piezoelectric and piezoelectric models.}\label{table:parameters}
\centering
\begin{tabular}{ccccc}
\hline
Parameter & PnC/Piezo PnC & LRM/Piezo LRM & IAM/Piezo IAM & Unit\cr
\hline
$a$ & 1 & 1 & 1 & m\cr
$\phi$ & --- & --- & 70 & degree\cr
$m_1$ & 0.1 & 0.1 & 0.01 & Kg\cr
$m_2$ & 0.3 & 0.3 & 0.3 & Kg\cr
$m_3$ & --- & --- & 0.09 & Kg\cr
$k_1$ & $4.7130\times10^5$ & $1.1844\times10^5$ & $0.9131\times10^5$ & Nm$^{-1}$\cr
$k_2$ & $1.5710\times10^5$ & $0.3948\times10^5$ & $3.0437\times10^4$ & Nm$^{-1}$\cr
$c_1$ & 3 & 3 & 3 & Nsm$^{-1}$\cr
$c_2$ & 3 & 3 & 3 & Nsm$^{-1}$\cr
$\alpha_1$ & 0.2605 & 0.1306 & 0.3626 & ---\cr
$\alpha_2$ & 0.3474 & 0.1741 & 0.1529 & ---\cr
$\beta_1$ & 3.7704 & 0.9475 & 7.3048 & ---\cr
$\beta_2$ & 1.6757 & 0.4211 & 0.3247 & ---\cr
${k_{\textrm{coeff}_1}}^2$ & 0.0430 & 0.1710 & 0.2218 & ---\cr
${k_{\textrm{coeff}_2}}^2$ & 0.0644 & 0.2565 & 0.3327 & ---\cr
\hline
\end{tabular}
\end{table}

The electrical parameters of the shunt circuits of the piezoelectric elements given in Table~\ref{table:parameters} are non-dimensionalized in the following manner:
\begin{subequations}\label{eq:non_dimensional_eqs}
\begin{align}
\alpha_l&=\bar{\omega}_lC_{p_l}R_l,\label{eq:alpha_eq}\\
\beta_l&={\bar{\omega}_l}^2C_{p_l}L_l,\label{eq:beta_eq}\\
{k_{\textrm{coeff}_l}}^2&=\frac{\theta_l^2}{k_lC_{p_l}},\label{eq:k_coeff_eq}
\end{align}
\end{subequations}
where $l=1,2$ is an index corresponding to the two degrees of freedom in each model, $\displaystyle\bar{\omega}_1=\sqrt{{k_1}/{m_1}}$, and $\displaystyle\bar{\omega}_2=\sqrt{{k_2}/{m_2}}$. The parameters $\alpha_l$, $\beta_l$, and ${k_{\textrm{coeff}_l}^2}$ are referred to as the non-dimensional resistor constant, non-dimensional inductor constant, and non-dimensional electromechanical coupling coefficient, respectively.

\subsection{Mathematical definitions of metadamping, energy-harvesting availability, and metaharvesting}\label{subsection:mathematical_definitions}

To quantify the emergent change in dissipation in the LRM and IAM chains compared to the statically equivalent PnC chain, an intrinsic wavenumber-dependent metric $ Z_{\textrm{emerg}_l}$ \cite{hussein2013metadamping} is used, which is defined as
\begin{equation}\label{eq:emergent_dissipation}
Z_{\textrm{emerg}_l}\vert_{*}(\mu)=\zeta_l\vert_{*}(\mu)-\zeta_l\vert_{\textrm{PnC}}(\mu)\ \ \ \ \ \ \ \begin{pmatrix}l=1, 2,\textrm{ or sum},\\ *=\textrm{LRM or IAM},\\ \mu\in[0,\pi] \end{pmatrix},
\end{equation}
where the subscript "emerg" stands for emergence; $l$ is an index referring to the acoustic branch $(l=1)$, optical branch $(l=2)$, or the summation $(l=\textrm{sum})$ of the two branches the subscript "$*$" is used to distinguish between the LRM and IAM, and $\mu$ is the dimensionless wavenumber defined as $\mu=\kappa a$. The quantity $ Z_{\textrm{emerg}_l}$ represents the difference between the damping ratio of the LRM or IAM and that of the statically equivalent reference PnC; hence, it is defined as the \textit{intrinsic wavenumber-dependent measure of the emergence of dissipation} in an LRM or IAL and illustrates the phenomenon of \textit{metadamping}. The wavenumber-dependent cumulative and total value of $Z_{\textrm{emerg}_l}$ are defined as
\begin{align}\label{eq:cumulative_emergent_dissipation}
Z_{\textrm{emerg}_l}^{\textrm{cum}}\vert_{*}(\mu)=\int_0^\mu Z_{\textrm{emerg}_l}\vert_{*}\ \textrm{d}\mu\ \ \ \ \ \ \ \begin{pmatrix}l=1, 2,\textrm{ or sum},\\ *=\textrm{LRM or IAM},\\ \mu\in[0,\pi]\end{pmatrix},
\end{align} 
and
\begin{equation}\label{eq:total_emergent_dissipation}
Z_{\textrm{emerg}_l}^{\textrm{tot}}\vert_{*}=Z_{\textrm{emerg}_l}^{\textrm{cum}}\vert_{*}(\pi)\ \ \ \ \ \ \ (l=1, 2,\textrm{ or sum};\ *=\textrm{LRM or IAM}),
\end{equation} 
respectively. The quantity $Z_{\textrm{emerg}_l}^{\textrm{cum}}$ represents the cumulatively integrated value of $Z_{\textrm{emerg}_l}$ over the irreducible Brillouin zone (IBZ), and $Z_{\textrm{emerg}_l}^{\textrm{tot}}$ is equal to $Z_{\textrm{emerg}_l}^{\textrm{cum}}$ evaluated at $\mu=\pi$ to give a single-valued total quantification of the overall level of metadamping pertaining to the acoustic branch, optical branch, or their summation for the LRM and IAM with respect to the reference PnC. Note that depending on the choice of the mechanical parameters, $Z_{\textrm{emerg}_l}, Z_{\textrm{emerg}_l}^{\textrm{cum}},\textrm{ and } Z_{\textrm{emerg}_l}^{\textrm{tot}}$ could be in the positive \cite{hussein2013metadamping,hussein2022metadamping} or negative \cite{hussein2022metadamping} regime.  

To quantify the changes in the damping ratio induced by the shunted piezoelectric elements in the piezoelectric media compared to their corresponding non-piezoelectric counterparts, the following intrinsic wavenumber-dependent metric is used~\cite{patrick2021brillouin}:
\begin{align}\label{eq:EHA}
Z_{\textrm{EHA}_l}\vert_{\textrm{P}}(\mu)=\zeta_l\vert_{\textrm{P}}(\mu)-\zeta_l\vert_{*}(\mu)\ \ \ \ \ \ \ \begin{pmatrix}
l=1, 2,\textrm{ or sum},\\ \textrm{P}=\textrm{Piezo PnC, Piezo LRM, or Piezo IAM},\\ *=\textrm{PnC, LRM, or IAM},\\ \mu\in[0,\pi]\end{pmatrix},  
\end{align}
where the subscripts $\textrm{P}$ and $*$ are used to distinguish among the three piezoelectric periodic media and the corresponding three non-piezoelectric periodic media, respectively. The quantity $Z_{\textrm{EHA}_l}$ represents the difference between the damping ratio of the piezoelectric periodic media and that of the corresponding reference non-piezoelectric periodic media, i.e., it is a measure of the useful dissipation arising purely due to the shunted piezoelectric elements and that is intrinsically available for harvesting by the same shunted piezoelectric elements considering that the raw dissipation, termed as "loss," is lost in the viscous damping dashpots with a prescribed damping. Hence, the measure $Z_{\textrm{EHA}_l}$ is defined as a \textit{wavenumber-dependent representation of the amount of useful dissipative energy intrinsically available for harvesting} in a piezoelectric periodic medium and illustrates the concept of \textit{EHA}. The wavenumber-dependent cumulative and total value of $Z_{\textrm{EHA}_l}$ are defined as
\begin{equation}\label{eq:cumulative_EHA}
Z_{\textrm{EHA}_l}^{\textrm{cum}}\vert_{\textrm{P}}(\mu)=\int_0^\mu Z_{\textrm{EHA}_l}\vert_{\textrm{P}}\ \textrm{d}\mu\ \ \ \ \ \ \ \begin{pmatrix}
l=1, 2,\textrm{ or sum},\\ \textrm{P}=\textrm{Piezo PnC, Piezo LRM, or Piezo IAM},\\ \mu\in[0,\pi]\end{pmatrix},
\end{equation} 
and
\begin{equation}\label{eq:total_EHA}
Z_{\textrm{EHA}_l}^{\textrm{tot}}\vert_{\textrm{P}}=Z_{\textrm{EHA}_l}^{\textrm{cum}}\vert_{\textrm{P}}(\pi)\ \ \ \ \ \ \ \begin{pmatrix}l=1, 2,\textrm{ or sum},\\ \textrm{P}=\textrm{Piezo PnC, Piezo LRM, or Piezo IAM}\end{pmatrix},
\end{equation} 
respectively. The quantity $Z_{\textrm{EHA}_l}^{\textrm{cum}}$ represents the cumulatively integrated value of $Z_{\textrm{EHA}_l}$ over the IBZ. Upon complete integration over the IBZ, $Z_{\textrm{EHA}_l}^{\textrm{tot}}$($Z_{\textrm{EHA}_l}^{\textrm{cum}}$ evaluated at $\mu=\pi$) is calculated to give a single-valued total quantification of the EHA pertaining to the acoustic branch, optical branch, or their summation for the piezoelectric periodic medium under consideration.
\begin{figure}[b]
\centering
\includegraphics{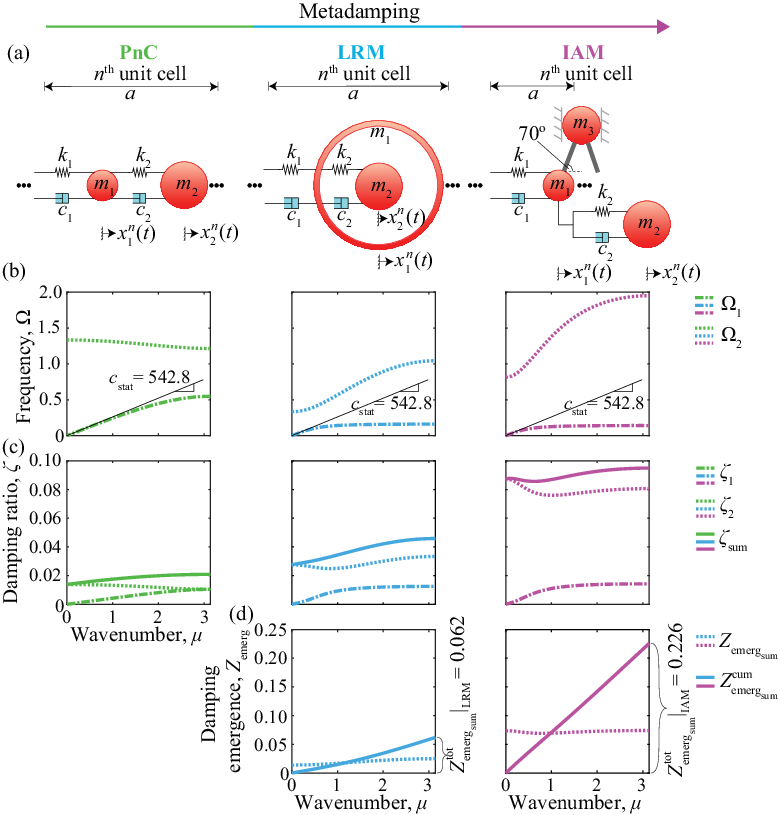}
\caption{Metamping: (a) Unit-cell schematics, (b) normalized damped-frequency band structures, and (c) damping-ratio diagrams for the statically equivalent PnC, LRM, and IAM ($\phi=70^\circ$) models. All models have the same number and type of damping dashpots with the same values of prescribed damping. Each damping-ratio diagram depicts the acoustic branch $\zeta_1$; the optical branch $\zeta_2$; and the summation of these two branches $\zeta_{\textrm{sum}}$. (d) Damping emergence in LRM and IAM relative to statically equivalent PnC. The damping emergence is calculated based on the summation of the two damping-ratio branches, $\zeta_{\textrm{sum}}$.}\label{fig:metadamping}
\end{figure}

To quantify the increase in the "useful dissipation" in the Piezo LRM and Piezo IAM relative to their statically equivalent Piezo PnC, an intrinsic wavenumber-dependent quantity $Z_{\textrm{EHA}\mathopen{\hphantom{g}}_l}^{\textrm{emerg}}$ is employed, defined as  
\begin{equation}\label{eq:emergent_EHA}
Z_{\textrm{EHA}_l}^{\textrm{emerg}}\vert_{\textrm{P}}(\mu)=Z_{\textrm{EHA}_l}\vert_{\textrm{P}}(\mu)-Z_{\textrm{EHA}_l}\vert_{\textrm{Piezo PnC}}(\mu)\ \ \ \ \ \ \ \begin{pmatrix}
l=1, 2,\textrm{ or sum};\\ \textrm{P}=\textrm{Piezo LRM, or Piezo IAM};\\ \mu\in[0,\pi]\end{pmatrix},
\end{equation}
where the subscript $\textrm{P}$ is used to distinguish between the Piezo LRM and the Piezo IAPM. Note that on the right-hand side in Eq. (\ref{eq:emergent_EHA}), the EHA (quantified by $Z_{\textrm{EHA}_l}$) is used instead of the damping ratio $\zeta_l$ as $Z_{\textrm{EHA}_l}$ excludes the raw (or lost) dissipation whereas $\zeta_l$ includes it. The quantity $Z_{\textrm{EHA}\mathopen{\hphantom{g}}_l}^{\textrm{emerg}}$ represents the difference between the EHA associated with the Piezo LRM or Piezo IAM and that associated with the statically equivalent reference Piezo PnC; hence, it is defined as the \textit{intrinsic wavenumber-dependent measure of the emergence of EHA} in a Piezo LRM or Piezo IAM. Thus the quantity $Z_{\textrm{EHA}\mathopen{\hphantom{g}}_l}^{\textrm{emerg}}$ illustrates the phenomenon of \textit{metaharvesting}. The cumulative and total value of $Z_{\textrm{EHA}\mathopen{\hphantom{g}}_l}^{\textrm{emerg}}$ are defined as
\begin{equation}\label{eq:cumulative_emergent_EHA}
Z_{\textrm{EHA}_l}^{\textrm{emerg}^{\textrm{cum}}}\vert_{\textrm{P}}(\mu)=\int_0^\mu Z_{\textrm{EHA}_l}^{\textrm{emerg}}\vert_{\textrm{P}}\ \textrm{d}\mu\ \ \ \ \ \ \ \begin{pmatrix}
l=1, 2,\textrm{ or sum},\\ \textrm{P}=\textrm{Piezo PnC, Piezo LRM, or Piezo IAM},\\ \mu\in[0,\pi]\end{pmatrix},
\end{equation} 
and
\begin{equation}\label{eq:total_emergent_EHA}
Z_{\textrm{EHA}_l}^{\textrm{emerg}^{\textrm{tot}}}\vert_{\textrm{P}}=Z_{\textrm{EHA}_l}^{\textrm{emerg}^{\textrm{cum}}}\vert_{\textrm{P}}(\pi)\ \ \ \ \ \ \ \begin{pmatrix}l=1, 2,\textrm{ or sum},\\ \textrm{P}=\textrm{Piezo PnC, Piezo LRM, or Piezo IAM}\end{pmatrix},
\end{equation}
respectively. The quantity $Z_{\textrm{EHA}_l}^{\textrm{emerg}^{\textrm{cum}}}$ represents the cumulatively integrated value of $Z_{\textrm{EHA}_l}^{\textrm{emerg}}$ over the IBZ. Upon complete integration over the IBZ, $Z_{\textrm{EHA}_l}^{\textrm{emerg}^{\textrm{tot}}}$ ($Z_{\textrm{EHA}_l}^{\textrm{emerg}^{\textrm{cum}}}$ evaluated at $\mu=\pi$) is calculated to give a single-valued total quantification of the emergent EHA or metaharvesting pertaining to the acoustic branch, optical branch, or their summation for the Piezo LRM and Piezo IAM with respect to the reference Piezo PnC.
\begin{figure}[b]
\centering
\includegraphics{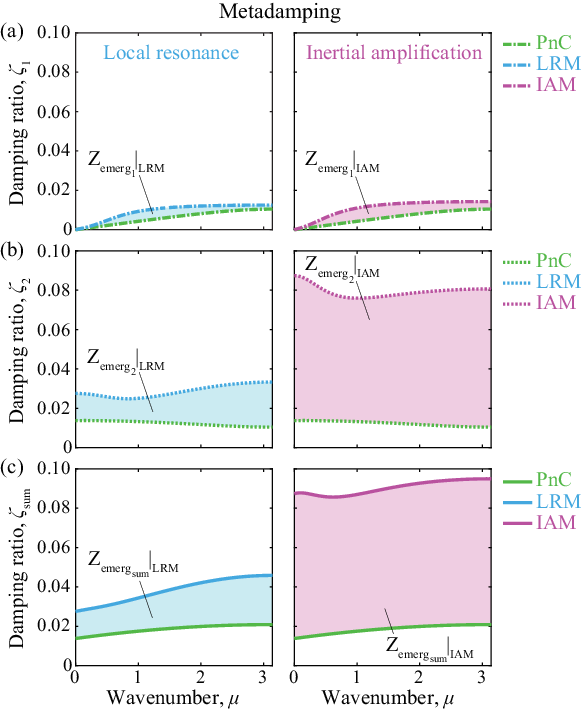}
\caption{Levels of metadamping (dissipation emergence) represented by shaded regions for LRM and IAM ($\phi=70^\circ$) relative to PnC. Plots show (a) acoustic damping-ratio branch $\zeta_1$; (b) optical damping-ratio branch $\zeta_2$; and (c) summation of the two aforementioned branches $\zeta_{\textrm{sum}}$ as a function of wavenumber.}
\label{fig:metadamping_summary}
\end{figure}

\subsection{Metadamping: Emergence in dissipation in locally resonant and inertially amplified metamaterials}\label{subsection:metadamping}

In this section, we examine the top horizontal arrow representing metadamping in the LRM and IAM models shown in Figs.~\ref{fig01} and~\ref{fig:unit_cell_schematics} using the parameters of Table~\ref{table:parameters}. Figure~\ref{fig:metadamping} depicts the unit-cell schematics of the PnC, LRM, and IAM, respectively, their non-dimensional damped-frequency band structures, their damping-ratio diagrams, and the quantity $Z_{\textrm{emerg}_l} (l=\textrm{sum})$ for the LRM and IAM models compared to the reference PnC model. In all the dispersion diagrams presented in this work, the damped frequencies associated with the acoustic and optical branches are plotted in non-dimensional form, i.e., $\Omega_l=\displaystyle{\omega_{d_l}}/{\bar{\omega_{1}}|_{\textrm{PnC}}}$, where $l=\textrm{1 (acoustic), 2 (optical)}$, and $\bar{\omega_{1}}|_{\textrm{PnC}}=\displaystyle\sqrt{{k_1|_{\textrm{PnC}}}/{m_1|_{\textrm{PnC}}}}$. All the non-dimensional damped-frequency band structures and damping-ratio diagrams are generated for $\mu\in[0,\pi]$, i.e., the IBZ. In Fig.~\ref{fig:metadamping}(b), the LRM and IAM each exhibit a narrower first (acoustic) transmission band and a wider second (optical) transmission band compared to the PnC. In Fig.~\ref{fig:metadamping}(c), the LRM exhibits higher damping ratios compared to the PnC and the IAM exhibits a significant further increase in the damping ratios, where the increase in both cases is stronger in the optical branch. Figure~\ref{fig:metadamping}(d) shows a positive level of metadamping (enhanced dissipation) in the LRM and an even higher level of metadamping in the IAM.

Figure \ref{fig:metadamping_summary} displays metadamping corresponding to the acoustic damping-ratio branch $\zeta_1$, optical damping-ratio branch $\zeta_2$, and the summation of the two branches $\zeta_{\textrm{sum}}$ as shaded regions highlighting the level of enhanced dissipation in each and also providing a visual contrast between the levels of enhanced dissipation for the LRM versus the IAM. The increase in the size of the shaded regions, $Z_{\textrm{emerg}_l}\ (l=1,\ 2,\textrm{ and }\textrm{sum})$, while progressing from the LRM to the IAM, indicates a significant increase in metadamping in the IAM. With a different choice of model paramters and/or moment-arm angle, the IAM chain could be designed to yield negative metadamping compared to the PnC~\cite{hussein2022metadamping}. Similarly, with a specific choice of model parameters and moment-arm angle, an IAM material may be realized that features a trade-off between temporal attenuation (metadamping) and spatial attenuation (band-gap effect)~\cite{hussein2022metadamping}. 

\begin{figure}[b]
\centering
\includegraphics{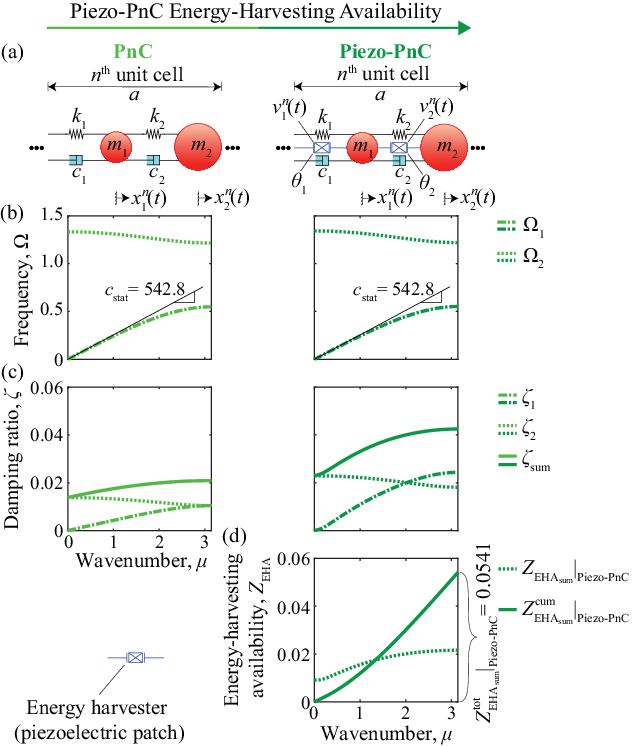}
\caption{Piezo-PnC energy-harvesting availability: (a) Unit-cell schematics, (b) normalized damped-frequency band structures, and (c) damping-ratio diagrams for the statically equivalent PnC and Piezo-PnC models, respectively. All models have the same number and type of damping dashpots with the same values of prescribed damping. Each damping-ratio diagram depicts the acoustic branch $\zeta_1$; the optical branch $\zeta_2$; and the summation of these two branches $\zeta_{\textrm{sum}}$. (d) EHA of Piezo-PnC, relative to statically equivalent PnC. The EHA measure is calculated based on the summation of the two damping-ratio branches $\zeta_{\textrm{sum}}$.}
\label{fig:PPnC_EHA}
\end{figure}
\subsection{Energy-harvesting availability: Intrinsic representation of useful dissipative energy available for harvesting}\label{subsection:EHAs}

\begin{figure}[!b]
\centering
\includegraphics{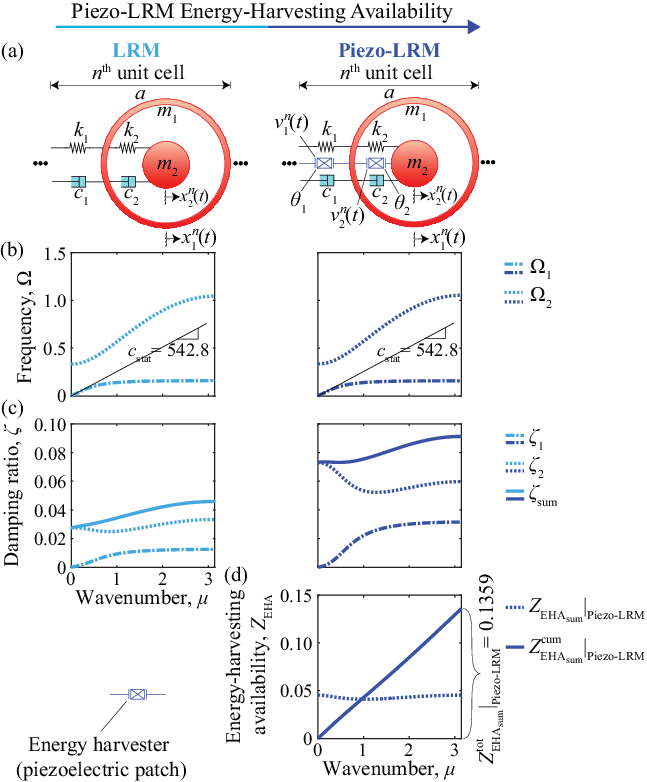}
\caption{Piezo-LRM energy-harvesting availability: (a) Unit-cell schematics, (b) normalized damped-frequency band structures, and (c) damping-ratio diagrams for the statically equivalent LRM and Piezo-LRM models, respectively. All models have the same number and type of damping dashpots with the same values of prescribed damping. Each damping-ratio diagram depicts the acoustic branch $\zeta_1$; the optical branch $\zeta_2$; and the summation of these two branches $\zeta_{\textrm{sum}}$. (d) EHA of Piezo-LRM, relative to statically equivalent LRM. The EHA measure is calculated based on the summation of the two damping-ratio branches $\zeta_{\textrm{sum}}$.}
\label{fig:LRPM_EHA}
\end{figure}

\begin{figure}[!h]
\centering
\includegraphics{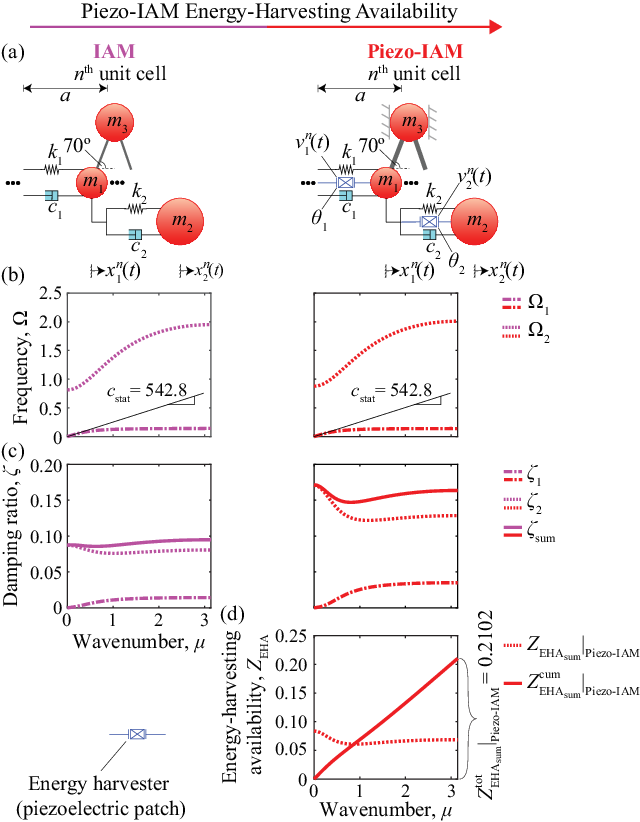}
\caption{Piezo-IAM energy-harvesting availability: (a) Unit-cell schematics, (b) normalized damped-frequency band structures, and (c) damping-ratio diagrams for the statically equivalent IAM ($\phi=70^\circ$) and Piezo-IAM ($\phi=70^\circ$) models, respectively. All models have the same number and type of damping dashpots with the same values of prescribed damping. Each damping-ratio diagram depicts the acoustic branch $\zeta_1$; the optical branch $\zeta_2$; and the summation of these two branches $\zeta_{\textrm{sum}}$. (d) EHA of Piezo-IAM, relative to statically equivalent IAM. The EHA measure is calculated based on the summation of the two damping-ratio branches, $\zeta_{\textrm{sum}}$.}
\label{fig:IALRPM_EHA}
\end{figure}

In this section, the vertical arrows assigned to EHA in the three piezoelectric periodic media in Fig.~\ref{fig:unit_cell_schematics} are studied and graphically illustrated. Figures~\ref{fig:PPnC_EHA},~\ref{fig:LRPM_EHA}, and~\ref{fig:IALRPM_EHA} depict the non-dimensional damped-frequency band structure, damping-ratio diagram, and EHA $Z_{\textrm{EHA}_l} (l=\textrm{sum})$ for the Piezo PnC, Piezo LRM, and Piezo IAM, respectively, in comparison to their non-piezoelectric versions.
In Figs.~\ref{fig:PPnC_EHA}-\ref{fig:IALRPM_EHA}, subfigures (b) show that the non-dimensional damped frequency band structures for the non-piezoelectric and piezoelectric periodic media are very similar.  Subfigures~\ref{fig:PPnC_EHA}(c),~\ref{fig:LRPM_EHA}(c), and~\ref{fig:IALRPM_EHA}(c) show that the piezoelectric periodic media exhibit significantly higher damping ratios relative to their non-piezoelectric counterpart; in subfigures (d) for the three piezoelectric periodic media, it can be observed that the total value of the EHA $Z_{\textrm{EHA}_{\textrm{sum}}}^{\textrm{tot}}$ corresponding to the summation of the two damping-ratio branches $\zeta_{\textrm{sum}}$ increases significantly while progressing from the Piezo PnC to Piezo LRM, and further to Piezo IAM.
\begin{figure}[!h]
\centering
\includegraphics{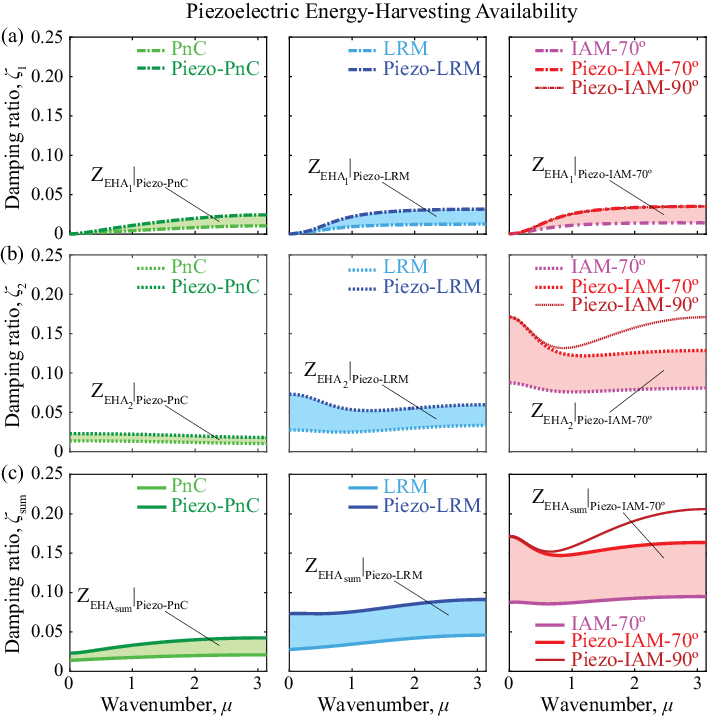}
\caption{Levels of EHA represented by shaded regions for Piezo-PnC, Piezo-LRM, and Piezo-IAM ($\phi=70^\circ$ and $\phi=90^\circ$) relative to PnC, LRM, and IAM ($\phi=70^\circ$), respectively. Plots show (a) acoustic damping-ratio branch $\zeta_1$; (b) optical damping-ratio branch $\zeta_2$; and (c) summation of the two aforementioned branches $\zeta_{\textrm{sum}}$ as a function of wavenumber.}
\label{fig:EHA_summary}
\end{figure}
We observe in Fig.~\ref{fig:PPnC_EHA}(c) that the damping ratio of the acoustic branch surpasses that of the optical branch at high wavenumbers for the Piezo PnC. In contrast, the optical branch damping ratio is consistently higher than that of the acoustic branch across the IBZ for both the Piezo LRM and Piezo IAM, as shown in Figs.~\ref{fig:LRPM_EHA}(c) and~\ref{fig:IALRPM_EHA}(c), respectively.~Figure~\ref{fig:EHA_summary} displays the EHA corresponding to the acoustic damping-ratio branch $\zeta_1$, optical damping-ratio branch $\zeta_2$, and the summation of the two branches $\zeta_{\textrm{sum}}$ as shaded regions to highlight the contrast among the three piezoelectric periodic media. The increase in the size of the shaded regions, quantified by $Z_{\textrm{EHA}_l}\ (l=1,\ 2,\textrm{ and }\textrm{sum})$, indicates a significant enhancement in the EHA while progressing from the Piezo PnC to Piezo LRM and an even greater enhancement while progressing from the Piezo PnC to Piezo IAM. When comparing the intrinsic EHA in the statically equivalent Piezo LRM and Piezo IAM, it is evident that inertial amplification causes a significant improvement particularly for the second branch. For the Piezo IAM, results are shown at two different inertial-amplifier angles: $\phi=70^\circ$ and $\phi=90^\circ$. An angle of $90^\circ$ is experimentally unrealisable but illustrates the maximum theoretical limit for the IAM. It is observed that $\zeta_1$ overlaps at $\phi=70^\circ$ and $\phi=90^\circ$ but $\zeta_2$ and $\zeta_{\textrm{sum}}$ are considerably higher at $\phi=90^\circ$.    

\subsection{Metaharvesting: Emergence of energy-harvesting availability in locally resonant and
inertially amplified metamaterials}
\label{subsection:metaharvesting}

\begin{figure}[!t]
\centering
\includegraphics{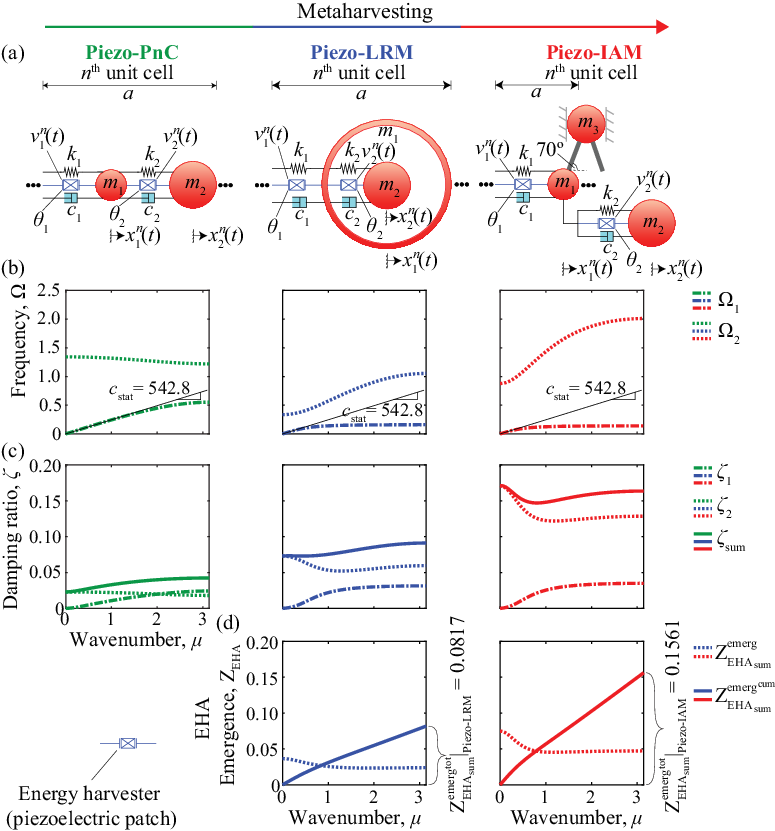}
\caption{Metaharvesting: (a) Unit-cell schematics, (b) normalized damped-frequency band structures, and (c) damping-ratio diagrams for the statically equivalent PnC, LRM, and IAM ($\phi=70^\circ$) models, respectively. All models have the same numbers and type of damping dashpots with the same values of prescribed damping. Each damping-ratio diagram depicts the acoustic branch $\zeta_1$; the optical branch $\zeta_2$; and the summation of these two branches $\zeta_{\textrm{sum}}$. (d) Emergent EHA in Piezo-LRM and Piezo-IAM, respectively, relative to statically equivalent Piezo-PnC. The EHA emergence is calculated based on the summation of the two damping-ratio branches $\zeta_{\textrm{sum}}$.}
\label{fig:metaharvesting}
\end{figure}

Now, the bottom horizontal arrow assigned to metaharvesting in the Piezo LRM and Piezo IAM in Fig.~\ref{fig:unit_cell_schematics} is examined and also graphically illustrated. Figure~\ref{fig:metaharvesting} shows the unit-cell schematics of the Piezo PnC, Piezo LRM, and Piezo IAM, their non-dimensional damped-frequency band structures, their damping-ratio diagrams, and the quantity $Z_{\textrm{EHA}\mathopen{\hphantom{g}}_l}^{\textrm{emerg}} (l=\textrm{sum})$ for the Piezo LRM and Piezo IAM. In Fig.~\ref{fig:metaharvesting}(c), it can be seen that the Piezo LRM exhibits higher damping ratios compared to the Piezo PnC and the Piezo IAM exhibits even higher damping ratios, where in both cases the increase is very significant in the optical branch. Thus, among the three statically equivalent piezoelectric periodic media, the Piezo IAM exhibits the highest overall dissipation. Figure~\ref{fig:metaharvesting}(d) shows an enhancement of EHA in the Piezo LRM and an even higher enhancement of EHA in the Piezo IAM.
\begin{figure}[!h]
\centering
\includegraphics{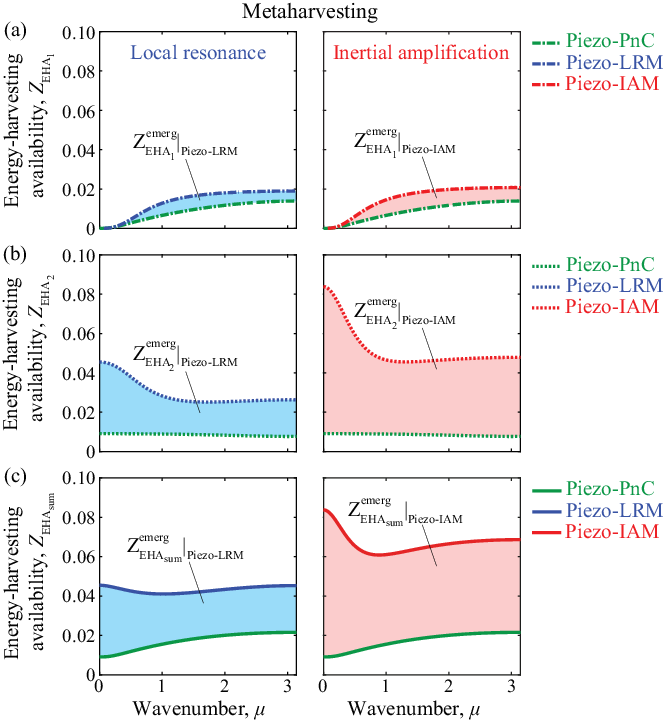}
\caption{Levels of metaharvesting (energy harvesting emergence) represented by shaded regions for Piezo-LRM and and Piezo-IAM ($\phi=70^\circ$) relative to Piezo PnC. Plots show (a) acoustic damping-ratio branch $\zeta_1$; (b) optical damping-ratio branch $\zeta_2$; and (c) summation of the two aforementioned branches $\zeta_{\textrm{sum}}$ as a function of wavenumber.}\label{fig:metaharvesting_summary}
\end{figure}
Figure~\ref{fig:metaharvesting_summary} illustrates the phenomenon of metaharvesting corresponding to the acoustic damping-ratio branch, optical damping-ratio branch, and the summation of the branches$-$each represented as shaded regions to highlight the contrast between the piezoelectric LRM and IAM. The increase in the size of the shaded regions, quantified by $Z_{\textrm{EHA}_l}^{\textrm{emerg}}\ (l=1,\ 2,\textrm{ and }\textrm{sum})$, indicates a significant enhancement in metaharvesting in the Piezo IAM compared to the Piezo LRM, which points towards better performance of an electromechanical finite structure made using an Piezo IAM material.

\section{Summary and conclusions}\label{section:summary_conclusions}

We have investigated three key types of piezoelectric phononic materials, namely, Piezo PnC, Piezo LRM, and Piezo IAM, and their corresponding non-piezoelectric counterparts. Using lumped-parameter modeling, all material systems were selected to exhibit the same long-wave static properties and feature the same level of prescribed damping in the dashpots. By applying Bloch's theorem for generalized free wave motion on all models, the wavenumber-dependent damped-frequency dispersion curves and corresponding damping ratio curves were calculated. Thus, all the analysis was done at the "material" level yielding intrinsic wave propagation properties that are independent of forcing, structural size, and boundary conditions.

A four-way "analysis square" framework as depicted in Figs.~\ref{fig01} and \ref{fig:unit_cell_schematics} was then followed, where the PnC and Piezo PnC are placed at the top-left and bottom-left corners, respectively, and the LRM/IAM and Piezo LRM/Piezo IAM are placed at the top-right and bottom-right corners, respectively. In this framework, a left-to-right arrow at the top edge of the square represents the phenomenon of metadamping$-$where the LRM and IAM materials were shown to exhibit enhanced raw dissipation compared to the PnC material. The top-to-bottom arrows along each side of the square provided a representation of the level of piezoelectric EHA, for the Piezo PnC on the left and the Piezo LRM and Piezo IAM on the right. Finally, a left-to-right arrow at the bottom edge of the square represents the phenomenon of metaharvesting$-$where the Piezo LRM and Piezo IAM materials were shown to exhibit enhanced piezoelectric EHA compared to the Piezo PnC. 
\begin{figure}[!t]
\centering
\includegraphics{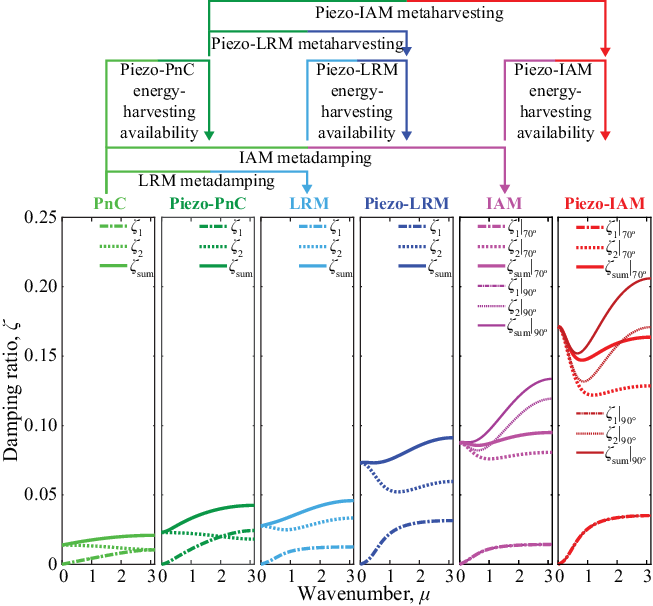}
\caption{Damping-ratio diagrams for PnC, Piezo-PnC, LRM, Piezo-LRM, IAM ($\phi=70^\circ$ and $\phi=90^\circ$), and Piezo-IAM ($\phi=70^\circ$ and $\phi=90^\circ$), respectively. Each damping-ratio diagram depicts the acoustic branch $\zeta_1$, the optical branch $\zeta_2$, and the summation of these two branches $\zeta_{\textrm{sum}}$. All models are statically equivalent and have the same type and level of prescribed damping. The various arrows at the top show the connections illustrated in Figs.~\ref{fig01} and~\ref{fig:unit_cell_schematics}.}
\label{fig:zeta_summary}
\end{figure}
\begin{figure}[!h]
\centering
\includegraphics{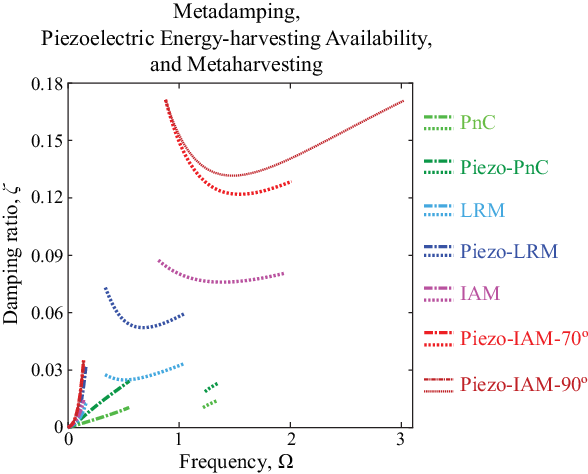}
\caption{Metadamping, piezoelectric energy harvesting, and metaharvesting: Single unified damping ratio-versus-frequency characterization diagram for all models considered in in Figs.~\ref{fig01} and \ref{fig:unit_cell_schematics}. The dash-dot lines show the relation for the acoustic branch, and the dotted lines show the relation for the optical branch. All models are statically equivalent and have the same type and level of prescribed damping.}
\label{fig:Omega_vs_zeta}
\end{figure}

Figure~\ref{fig:zeta_summary} shows the damping-ratio diagrams of all the non-piezoelectric and piezoelectric materials investigated. Upon comparison between the various subfigures, the levels of metadamping, EHA, and metaharvesting are directly determined. Fig.~\ref{fig:Omega_vs_zeta} provides another graphical summary, compactly showing the relation between the damped frequencies $\Omega_1$ and $\Omega_2$ with their corresponding damping ratios $\zeta_1$ and $\zeta_2$, respectively, for all models. Once again, the levels of metadamping, EHA, and metaharvesting are directly observed when comparisons are made following the analysis arrows shown in Figs.~\ref{fig01} and \ref{fig:unit_cell_schematics}.

The results show that the intrinsic EHA may be enhanced by utilizing the local-resonance mechanism and even further enhanced by employing the inertial-amplification mechanism. This emergent enhancement of EHA, termed metaharvesting, is formally characterized by comparing with a statically equivalent Piezo PnC with the same level of prescribed damping. These findings formulate a pathway towards design of architectured piezoelectric materials with superior energy harvesting capacity guided by fundamental phononic principles. Once a superior piezoelectric unit-cell design is identified in the current form of a simplified lumped parameter model, the next step is to convert it to a physically realizable material configuration, such as, for example, rod or beam components with extruded pillars for local resonance~\cite{badreddine2012enlargement,bilal2013trampoline} or attached stiff levered substructures for inertial amplification~\cite{yilmaz2007phononic,frandsen2016inertial}. 

\bibliography{References_metaharvesting}


\end{document}